\documentclass[twocolumn,aps,floatfix]{revtex4-1}
\usepackage{amssymb}
\usepackage{graphicx}
\usepackage{amsmath}
\usepackage{pstricks}

\begin{document}
\title{Two component Bose-Hubbard model  with higher angular momentum states}
\author{\mbox{Joanna Pietraszewicz$^{1}$, Tomasz Sowi\'nski$^{1,3}$, Miros{\l}aw Brewczyk$^2$,} \\ Jakub Zakrzewski$^{5}$, \mbox{Maciej Lewenstein$^{3,4}$, and Mariusz Gajda$^{1}$}}
\affiliation{
\mbox{$^1$Institute of Physics Polish Academy of Sciences, Al. Lotnik\'ow 32/46, 02-668 Warszawa, Poland}\\
\mbox{$^2$Wydzia{\l} Fizyki, Uniwersytet w Bia\l ymstoku, ul. Lipowa 41, 15-424 Bia\l ystok, Poland} \\
\mbox{$^3$ICFO - Institut de Ci\`ences Fot\`oniques, Parc Mediterrani de la Tecnologia, E-08860 Castelldefels, Barcelona, Spain} \\
\mbox{$^4$ ICREA - Instituci\'o Catalana de Recerca i Estudis Avan\c cats, 08010 Barcelona, Spain}
\mbox{$^5$Instytut Fizyki im. Mariana Smoluchowskiego, Uniwersytet Jagiello\'nski, ul. Reymonta 4, 30-059 Krak\'ow, Poland}
}
\date{\today}

\begin{abstract}
Bose-Hubbard Hamiltonian of cold two component Bose gas
of spinor Chromium atoms is studied. Dipolar interactions of magnetic moments
while tuned resonantly by ultralow magnetic field can lead to a 
transfer of atoms from the ground to  
excited Wannier states with a  non vanishing angular orbital momentum. 
Hence we propose the way of creating of $P_x+i P_y$ orbital superfluid. 
The spin introduces an additional degree of control and
leads to a variety of different stable phases of the system. The Mott insulator 
of atoms in a superposition of the ground and vortex Wannier states as well
as a superposition of the Mott insulator with orbital superfluid are predicted.
\end{abstract}

\maketitle

\section{Introduction}

Ultracold atoms provide a playground for mimicking condensed
matter and studying novel quantum many-body
phenomena \cite{Bloch,LewenR}. Recently, there has been
particularly impressive progress in two areas of physics of
ultracold atoms: the area of ultracold dipolar
gases \cite{Baranov,Lahaye,Santos,Sengstock} and the physics of orbital
lattices \cite{LewenNP}. In this paper we combine these two
areas and explore the effect of two-body dipolar interactions of
magnetic atomic moments in a lattice potential. We study dipolar gases
in their full complexity including spin as a dynamical variable
(as opposed to be a conserved quantity), and magnetic dipolar interactions
coupling different orbital states
of involved magnetic components. That introduces additional
physical processes into  play  and new degrees of 
control to the standard Bose-Hubbard model.

Spinor gases in a lattice have been studied in the context
of Mott insulator (MI)- superfluid (SF) transition \cite{Kimura}. 
In general  dipolar interactions
lead directly to the dynamics of spin degree of freedom but up till now in lattice systems
this phenomenon was neglected, i.e. it was assumed that spin is frozen \cite{Lahaye}.
In such situations electric and magnetic dipoles are practically equivalent -- they
introduce long range correlations. On the other hand it is known from studies
of gases confined in harmonic traps in the mean field limit \cite{Baranov,Lahaye} that 
taking the dynamics of spin into account may modify  properties of the ground state of the system.

Spin dynamics may result from contact or dipolar
interactions. In the former case the total spin of interacting atoms
remains unchanged (magnetization of the sample is constant). Qualitatively different
phenomena take place when spin dynamics is triggered by the dipolar forces. The atomic magnetic
moment originates from the 
spin which contributes to the total angular momentum of the system. When magnetic dipole
changes due to the dipolar interactions, its variation must
be accompanied by corresponding dynamics of the orbital angular momentum.
Magnetic interactions can lead to a transfer of angular momentum from spin to orbital
degrees of freedom.  This phenomenon, discovered in
ferromagnetic solid samples, is known as Einstein-de Haas effect \cite{Gawryluk,Ueda,Swistak1,Swistak_plakietki}.
Not a long range character of magnetic dipolar
interactions but rather their relation to the angular momentum
plays a crucial role in this phenomenon. This makes
a fundamental difference between magnetic and electric dipoles.

The main issue of our study is to account for the spin
degree of freedom in the lattice environment.
Spin flipping processes in the lattice could lead
to an appearance of the orbital $P_x+iP_y$ superfluid. Recently
orbital superfluids were created in experiment \cite{Hammer}.
The authors utilized a resonant tunneling in a particularly designed
lattice potential. 

In this paper we show another way
of creating orbital superfluid by means of the resonant Einstein-de Haas effect.
The atom which flips its spin has to gain
some additional kinetic energy necessary to support its rotation.
This energy is typically much larger than the energy of dipolar
interactions and conservation of energy strongly suppresses
the spin dynamics. The transfer of atoms between  two spinor components can be enhanced
by tuning energies of states involved via Zeeman effect 
 \cite{Swistak1}. We extend this idea to lattice gases.
Dipolar effects significantly modify the MI-SF transition 
lead to new phases of the system with quantized vortices 
in MI or/and SF regimes.

The paper is oranized as follows: in Section II we introduce the two component
Bose-Hubbard model with dipolar interactions coupling different Wannier states,
in Section III we present a phase diagram for the system while in Section IV
we discusse validity and limiatations of the model.
 
\section{The model}

We assume that Cr atoms are in a 2D optical square lattice.
To fix the parameters we consider a realistic situation of
the lattice described by the periodic  potential $V_0 [\sin^2(2\pi x)+\sin^2(2 \pi y)]$
Here $\lambda=523\,\mathrm{nm}$ is the wavelength
of light beams creating the lattice and $V_0$ is the barrier height. 
A characteristic energy of the problem, i.e. the  recoil energy is
$E_r=\hbar^2 (2 \pi)^2/(2m \lambda^2)$. 
We express all energies and lengths
in units of $E_r$ and $\lambda$ respectively.  
Confinement along the $z$ direction is provided by
a harmonic potential $m \omega_z^2 z^2/2$ of frequency $\hbar \omega_z=16E_r$.
At each lattice site we choose 
two wave functions centered at the given site $(x_i,y_i)$ to form
a single particle basis of the two component system. The basis
allows to account for the resonant transfer of atoms between
$m_S=3$, $l=0$ and $m_S=2$ and $l=1$ states in the presence of
magnetic field aligned along the z-axis.
The lowest energy state $\psi_a (x,y,z) \sim {\cal
W}_0(x){\cal W}_0(y)\exp(-z^2\omega_z/2)$ is effectively 
coupled to the excited state with
one quantum of orbital angular momentum $\psi_b (x,y,z) \sim
\left[{\cal W}_1(x){\cal W}_0(y) + i{\cal W}_0(x){\cal
W}_1(y)\right]\exp(-z^2\omega_z/2)$. The state is a single
site analogue of 
a hamonic oscillator state $\sim (x+iy) \exp[-(x^2+y^2)/2-z^2 \omega_z/2]$. 
${\cal W}_0(x)$ and ${\cal
W}_1(x)$ are the ground and the first excited Wannier states in a
1D periodic potential of the form  $V_0 \sin^2(2\pi x)$.
Single particle energies of the two essential states
are denoted by $E_a$ and $E_b$ respectively. 

Limiting the subspace of essential states is a crucial approximation
in our study. It is possible only
due to a weakness of dipolar interactions.
In fact there are several channels of binary dipolar collisions
leading to different excited Wannier states.
However, we can choose the desired channel
by a proper adjustment of the resonant external magnetic field \cite{Swistak1}.
Typically the energy difference between
atoms in the ground and in the excited Wannier states
is much larger then dipolar energy which is 
the smallest energy scale in the problem (except vanishing
tunnelings case), $E_{dip}=10^{-4} E_r \ll E_b-E_a \sim E_r$.
However, at resonant magnetic field $B_0$, $E_a-g\mu_B B_0 = E_b$, the two energies are equal and
the spin transfer between the components becomes efficient
on a typical time scale $\hbar/E_{dip} \simeq 10^{-2}$s.
Here $\mu_B$ is the Bohr magneton and $g=2$ is the Lande factor. Only then the system
can dynamically redistribute particles between the two components
without violating energy conservation.
A characteristic width of the resonances is small \cite{Brewczyk},
of the order of $E_{dip} \approx g \mu_B B$, i.e. $ B \approx 100\mu$G.
We assume that no other states can be effectively coupled (see a more detailed discussion
of the validity of this model in Section IV).

In effect a two-component system is realized with
$a$-component corresponding to atoms in $m_S=3$ and $l=0$ state
while atoms in $b$-component have $m_S=2$, $l=1$. Single site basis
states are $|n_a,n_b \rangle$, where $n_{c}$ is a number
of atoms in $c$-component ($c=a,b$). The Hamiltonian of the system is:
\begin{eqnarray}
H &=&  \sum_i \left[ (E_a-g\mu_B B)\,a_i^\dagger a_i + E_b\,b_i^\dagger b_i  +  U_{ab}\,a_i^\dagger b_i^\dagger a_ib_i \right. \nonumber \\
&& \left. + \frac{U_a}{2}a_i^{\dagger 2} {a_i}^2  + \frac{U_b}{2}{b_i}^{\dagger 2} {b_i}^2   + 
D({b_i}^{\dagger 2}  {a_i}^2 + a_i^{\dagger 2} {b_i}^2) \right]\nonumber \\
&& -\sum_{\langle i,j\rangle}\left[ J_a\, a_i^\dagger a_j  + J_b\, b_i^\dagger b_j \right].
\label{BH_Hamiltonian}
\end{eqnarray}
The parameters  depend only on lattice height $V_0$
and confining frequency $\omega_z$ in the z-direction. 
$U_a, U_b, U_{ab}$ are the contact interaction energies plus
the part of dipolar energy which has the same form as corresponding contact term,
$D$ is the on-site dipolar coupling of the two components, while $J_a$
and $J_b$ are tunneling energies.
The Hamiltonian (\ref{BH_Hamiltonian}) is an interesting modification of the
standard Bose-Hubbard model.

The on-site contact interactions
$U_{a}, U_b$, and $U_{ab}$ cannot change a total spin \cite{Ueda_review,bruno}. 
Dipolar two body interactions are much smaller than the contact ones;
we keep only those dipolar terms which lead to a spin dynamics.
Moreover, only on-site dipolar effects
are accounted for in the Hamiltonian (\ref{BH_Hamiltonian}).
Dipolar potential, although long
range, is so weak that we can ignore dipole-dipole interactions
between atoms at neighboring sites in the considered range of small
tunnelings. 

Unlike tunneling between ground Wannier states $J_a$, the
tunneling energy $J_b$ of the excited state is negative because
the wave function the $\psi_b(x,y,z)$ is antisymmetric in $x$ and $y$. 
Therefore the state with `antifferomagnetic' order of phases 
between neighboring sites has lower energy than the state where phases
of the exited Wannier functions are the same. 
For the opposite on-site phases of the excited Wannier states
both $J_a$ and $J_b$ are positive. This case is considered here.

\section{Phase diagram of the model}

We limit our study to a small occupation of a lattice site: not more than one
particle per single site on average. The resonant magnetic fields
equilibrates single particle energies of states $|1,0\rangle$ and
$|0,1\rangle$, i.e. $E_b=E_a-g\mu_B B_0$.
$E_a$ and $E_b$ depend on the lattice height thus the resonant magnetic
field varies with $V_0$, $B_0=B_0(V_0)$. 

Even with a single particle per site the dipolar
interactions couple ground and excited Wannier states due to
the tunneling in a higher order process. 
The transfer between $|1,0\rangle$ and  $|0,1\rangle$ states is a sequence of:
adding an atom to the $a$-component at a given single site $|1,0\rangle \rightarrow
|2,0\rangle$ via tunneling, followed by the dipolar transfer of both $a$-species
atoms to the excited Wannier state $|2,0\rangle \rightarrow
|0,2\rangle$, and finally the tunneling which removes one
$b$-component atom from the site $|0,2\rangle \rightarrow
|0,1\rangle$. The two considered states are therefore
coupled provided that tunneling is nonzero.

Now, following the standard mean field approach of Fisher et al.\cite{Fisher}
we find thermodynamically stable phases of the system in the choosen subspace.
The Hamiltonian (\ref{BH_Hamiltonian}) is translationally invariant, we assume the
same property is enjoyed by the lowest energy state. Introducing
superfluid order parameters for both components: $\phi_{(a)}=\langle
a_i \rangle$ and  $\phi_{(b)}= \langle b_i \rangle$ as well as
the chemical potential $\mu$, the Hamiltonian of the system 
can be approximated by a sum of single site Hamiltonians $H_0+H_I$
\begin{eqnarray}
\label{H0}
H_{0} &=& -\mu(a^\dagger a+b^\dagger b)  + \frac{1}{2}U_a a^\dagger a^\dagger a a +\frac{1}{2} U_b b^\dagger b^\dagger b b \nonumber \\
&&+ U_{ab} a^\dagger b^\dagger ab + D(b^\dagger b^\dagger aa + a^\dagger a^\dagger b b),\\
H_{I} &=& -zJ_a {\phi}^*_{(a)} a  - zJ_b {\phi}^*_{(b)} b +h.c.
\label{hop}
\end{eqnarray}
Notice we skipped indices enumerating sites. In (\ref{hop})
$z$ is a number of neighbors and depends on the lattice
geometry. For a 2D square lattice $z=4$. Hamiltonian $H_0+H_I$
does not conserve number of particles: it describes 
a single site coupled to a particle reservoir.  
Order parameters $\phi_{(a)}$ and $\phi_{(b)}$ vanish in the MI phase
and hopping of atoms is suppressed. Only in the SF regime number of
particles per site can fluctuate. Close to  the boundary, on the
SF side, $\phi_{(a)}$ and $\phi_{(b)}$ can be treated as small
parameters of the perturbation theory.

\begin{figure}
\includegraphics[scale=0.7]{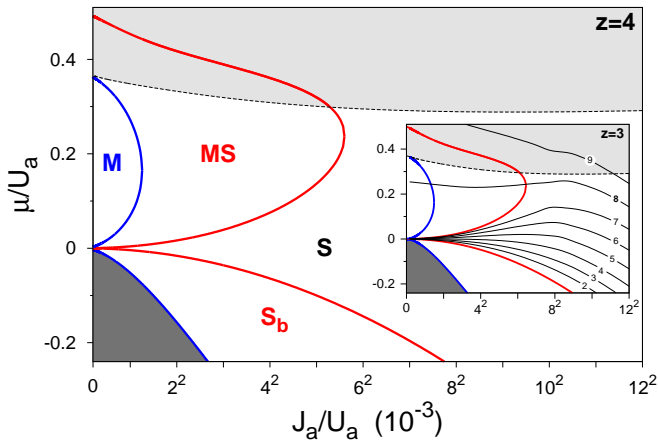}
\caption{Phase diagram for 2D square lattice at the resoance, $z=4$. The
regions are: $M$ -- Mott insulator with one particle in equal superposition
od $a$ and $b$ states, $MS$ -- superfluid in  $a$ and $b$ components
($b$-dominated) and Mott insulator in the orthogonal supperposition,  
$S$ -- superfluid phase of superposition of $a$ and $b$ components,
$S_b$ -- superfluid in the $b$-component.
In the inset the diagram for $z=3$ together with chemical potential $\mu(N)$
for a given number of particles obtained from the exact diagonalization. 
The lines, from bottom to top correspond to occupation equal to 
$N=2, \hdots, 9$ as indicated. For $\mu> U_b$ (light grey region) the ground state 
of the system is a two particle state, therefore in this regime, the phases shown
are thermodynamically unstable. They are stable, however, with respect
to one particle hopping.}\label{fig:lob2}
\end{figure}

The single site ground state becomes unstable if the mean field $\phi_{(a)}$ 
or $\phi_{(b)}$ are different than zero. The mean fields can be obtained numerically from the
self-consistency condition:
\begin{align}
\phi_{(c)}=\lim_{\beta\rightarrow \infty}\mathrm{Tr}\left[c\mathrm{e}^{-\beta( H_0+H_I)}\right]/Z({\beta}),
\label{sfcond}
\end{align}
where $c=a,b$.
In the lowest order of the perturbation in
the order parameters, the set of equations (\ref{sfcond})
becomes linear and homogeneous. Vanishing of its determinant is a necessary condition
for nonzero solutions for $\phi_{(c)}$. This condition determines lobs shown in 
Fig.~\ref{fig:lob2}. 

In the low temperature limit ($\beta \rightarrow
\infty$) the partition function reduces to a
single lowest energy state contribution $Z(\beta)=\mathrm{e}^{-\beta E_0}$. 
The energy $E_0$ depends on the chemical potential $\mu$.
Moreover, for $\mu <U_{b}<U_{a}$ the only contribution to Eq.(\ref{sfcond}) comes 
from eigenstates of the Hamiltonian  with  
zero, one and two particles. Our analysis is limitted to this case only.

For negative chemical potential, $\mu<0$, 
the single site ground state is $|0,0\rangle$ vacuum state (dark grey region in Fig.~\ref{fig:lob2}).
With increasing tunneling (and fixed $\mu$) particles appear in the superfluid vortex $b$-phase ($S_b$).  
Only at larger tunnelings  some atoms do appear in the $a$-component and both:
`standard' and $P_x+iP_y$ orbital superfluids coexist ($S$). 

Situation becomes more complicated for larger chemical potential $0<\mu < U_{b}$.
At the resonance, $B=B_0$, the
ground state is degenerate if tunneling is neglected:  the states $|1,0\rangle$ and $|0,1\rangle$ 
have the same energy, $E_0=-\mu$.
The degeneracy is lifted via tunneling
in the second order of the perturbation. In addition 
a position of the resonance is shifted towards smaller magnetic field values.
Analysis of the effective Hamiltonian (compare \cite{Grass}) indicates that in the resonant region the 
single site ground state is a supperposition of both components
$|g\rangle= \alpha_1 |1,0\rangle- \alpha_2 |0,1\rangle$.
Exactly at  resonance $\alpha_1=\alpha_2 =1/\sqrt{2}$. 
While crossing the resonance the ground state switches from $|1,0\rangle$ 
to $|0,1\rangle$.  The width of the resonance $\Delta B $ can be 
estimated perturbatively to be $g\mu_B |\Delta B| \approx 10^{-6} E_r$ 
for $V_0=25 E_r$ while  for lower barriers,
$V_0 = 10 E_r$, the resonant region is broader $g\mu_B |\Delta B| \approx 10^{-3} E_r$.
Due to its small width  the resonance  can be hardly accesible particularly for small tunnelings. 
Away from the resonance the standard phase diagrams for
$a$ or $b$ component emerge.

In Fig.~\ref{fig:lob2} we show regions of stability of different
possible phases of the system at resonance i.e when  
$|g\rangle= (|1,0\rangle - |0,1\rangle)/\sqrt{2}$. 
For small tunnelings the system is in the Mott insulating phase (M) with one atom per site. 
Every atom is in the {\it superposition} of the ground and the vortex Wannier state. At the blue line,
the border of (M) lobe, Eqs.(\ref{sfcond}) allow for nonzero solutions for  $\phi_{(a)}$
and $\phi_{(b)}$.  Eqs.(\ref{sfcond}) become diagonal if $H_I$ is
expressed in terms of bosonic operators $A^{\dagger}=(\kappa_a a^{\dagger} + \kappa_b b^{\dagger})$
and $B^{\dagger}=(-\kappa_b a^{\dagger} + \kappa_a b^{\dagger})$ where $\kappa_a^2+\kappa_b^2=1$ and
both coefitients of the superposition depend on the tunellings $J_a$ and $J_b$.
The operators create an atom in two orthogonal superpositions of $a$ and $b$ states.
At the border of the Mott phase (M) 
the mean value of the operator $B$ is different from zero and a nonvanishing 
superfluid component,  $\Psi_B = -\kappa_b \phi_{(a)} + \kappa_a  \phi_{(b)}$, 
appears in the (MS) region. Our numerical results show that 
$\kappa_a \simeq - 0.99$ and the ratio $(\kappa_b/\kappa_a)^2 \simeq 0.02$ is small  
at the edge of stability of the Mott insulator. Therefore $B^{\dagger} \simeq b^{\dagger}$,
i.e. the superfluid $\Psi_B$ is dominated by the orbital $b$-component. The 
mean field corresponding to the $A^{\dagger} \simeq a^{\dagger}$ operator is zero
in the discussed region. The system is therefore in  equal {\it superposition} of the Mott insulating 
and superfluid phases. The Mott phase is 
dominated by the $a$-component and the superfluid phase is overwhelmed with 
the $b$-species. Both components, however, contain a small minority of remaining species.

At larger tunneling the system undergoes another phase transition as Eqs.(\ref{sfcond}) allow for
another nonzero mean field. Now the mean value of $A$ departs form zero defining the border of the `bigger'
lob. Mott component of the ground Wannier state becomes unstable.
The additional mean field $\Psi_A = \kappa_a \phi_{(a)} + \kappa_b \phi_{(b)}$ appears in the (S) region.
Again   $\kappa_a \simeq 0.97$ and the maximal value of  
$(\kappa_b/\kappa_a)^2 \simeq 0.06$ is small. The $a$-species 
dominate the $\Psi_A$ superfluid component. Both $\Psi_A$ and $\Psi_B$ superfluids exist in the (S) region.   

All the above findings are supported by direct inspection of the 
true many body ground state obtained by exact digonalization of 
the many body Hamiltonian in a small $2\times 4$ rectangular plaquette
with periodic boundary conditions 
for total number of particles $N=1, \ldots, 10$.
Note that each site has three neighbors, $z=3$, in this case. 
Resonance condition is reached by finding the magnetic field for which both $a$ and $b$ species
are {\it equaly populated}. Calculations for 
$z=4$ require much larger number of sites and are numerically unreachable.
In the inset of Fig.~\ref{fig:lob2} we compare the exact results 
with the mean field ones but for $z=3$. 
The lines in the inset correspond to the constant number
of particles per site obtained from the relation
$\mu(N) = \left[E_0(N+1)-E_0(N-1)\right]/2$. They allow
to trace the phases the system enters while adiabatically
changing the tunneling at fixed particle number. 
The (M) and (MS) phases can be reached with one particle per site only
(8 particles in the plaquette).
Direct inspection of a structure of the many body ground state 
fully confirms the stable phases of the system described above. In particular
the ground state in the (MS) region can be approximated (with the accuracy of about 4\%)
by $\frac{1}{\sqrt{2}}[\Pi a_i^{\dagger}-\frac{1}{\sqrt{N!}}(\frac{1}{\sqrt{N}}\sum b_i^{\dagger})^N]|\Omega\rangle$,
where $|\Omega\rangle$ is the vaccum state.

\begin{figure}
\includegraphics[scale=0.7]{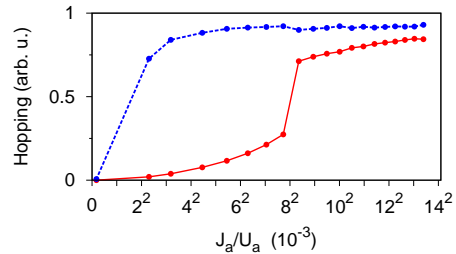}
\caption{Hopping for the lowest energy state
in a $2 \times 4$ plaquette obtained from the exact diagonalization. Upper line -- $b$ component, lower line -- $a$ component.}
\label{fig:exact}
\end{figure}

In addition we calculated a hopping, i.e. the mean values of the following hopping operators :
$h_{a} =\sum_{\langle j\rangle}
\langle a^\dagger_j a_i \rangle$ and $h_b=\sum_{\langle j
\rangle} \langle b^\dagger_j b_i \rangle$. 
These operators annihilate a particle at a given site and put it in a neighboring site.
They might be viewed  as number conserving analogons of the mean fields $\phi_{(a)}$ and $\phi_{(b)}$. 
In Fig.~\ref{fig:exact} we show the hopping for the case of one particle per site.
For large tunnelings both $a$ and $b$ hopping are large -- the components are 
in the superfluid phase. Entering the MS phase, $J_a/U_a \simeq 0.064$, the hopping
of $a$-component rapidly falls down while hopping of $b$-atoms remains big  -- the system
enters $a$-component dominated Mott insulator superimposed with $b$-component dominated superfluid.
At $J_a/U_a \simeq 0.002$ both hoppings tend to zero -- the system enters the Mott phase
with equal occupation of both species. This confirms results based on the Fisher method. 

\section{Validity of the model}

Finally let us discuss possible limitations of the validity of the model discussed above. 
As we study a stability of the Mott phase we consider the case of deep optical lattices
where tunneling is a small perturbation only.  It is very natural to assume that 
dipolar interactions couple the ground Wannier state to the orbital state 
at each lattice site, and the system posseses the  translational symmetry. 
Moreover, we have assumed that {\it locally} the potential at a given site
has almost perfect axial symmetry with respect to the site center.
Therefore, the local site Hamiltonian preservs projection of the total angular
momentum, and the only state coupled to the ground Wannier one is of the type 
$\sim (x+iy)$, where $x$ and $y$  are measured  with respect to the site center. 
This state is the eigenstate of the projection of the orbital angular momentum on the $z$-axis.  

Three comments are in order.

\subsection{ Role of anharmonicity.}

Due to high selectivity of magnetic
resonances we have a freedom of choosing a given channel of dipolar collision by a proper adjustment of the external magnetic field.
In particular, we study the channel where the $z$-comonent of the relative orbital angular momentum
of interacting  particles changes by two quanta, $\Delta L_z=2$.  Assuming that 
each of two colliding atoms are initially in the spherically symmetric ground state, the lowest energy 
final state of the two atoms has a form $|{\rm vortex}\rangle \sim (x_1+iy_1)^2+(x_2+iy_2)^2-2(x_1+iy_1)(x_2+iy_2)$.
Note that in the harmonic trap of radial frequency $\omega$ the state  $|{\rm vortex}\rangle$ corresponds to a superposition
of  two states: $|{\rm v_2}\rangle\sim (x_1+iy_1)^2+(x_2+iy_2)^2$ and $|{\rm v_1}\rangle\sim (x_1+iy_1)(x_2+iy_2)$, 
where $x_i,\  y_i$ are particles coordinates.
For $|{\rm v_2}\rangle$ one of colliding atoms aquires two quanta of rotation 
while the second atom remains in the spatial ground state, i.e. the energy of $|{\rm v_2}\rangle$ is $E_2 = 2\hbar \omega$. 
On the other hand $|{\rm v_1}\rangle$ represents the situation where each of two atoms gets one quantum
of rotation resulting in the total energy $2E_1=2\hbar \omega$. 
Evidently for equally spaced harmonic  energy levels  both states are degenerate, $E_2=2E_1$
and both the conservation of angular momentum and the conservation of energy can be satisfied.

The situation becomes different in the optical lattice because of anharmonicity of the lattice potential.
The state $|{\rm v_2}\rangle$ has energy of the second Wannier state, $E_2$.
This energy is smaller than twice the energy of the first excited Wannier $2 E_1$ of the state $|{\rm v_1}\rangle$.
Even for high barriers, i.e. $V_0=40E_r$ the energy splitting $2E_1-E_2=10^{-1} E_r$
is significantly larger then dipolar energy of Cr atoms, $E_{dip}=10^{-4}E_r$. Therefore, by means of magnetic field tuning
one may select
the resonant transfer of atoms due to dipolar interactions bringing both interacting particles to the state 
with one quantum of rotation while making the transfer to  the second Wannier state nonresonant (and not efficient).
This is a situation considered in the present paper. Note that a proper adjustment of the magnetic field may
make the excitation of $D$ orbital $|{\rm v_2}\rangle$ resonant - the situation not considered here.  
Anharmonicity of the lattice potential, although small, plays thus  an important role.

\subsection{Vorticity versus tunneling}

The second issue is related to the tunneling of vortex-like states. As the lattice states have $C_4$ 
symmetry the angular momentum need not  be conserved in the tunneling. With our choice of alternating
phases of the excited Wannier states the tunneling coefficient in the excited band is positive, $J_b>0$.
Therefore tunneling of the right handed vortex $\sim (x+iy)$ to the vortex
of the same vorticity at neighbouring site is equal to $t_R=J_b+J_a$ and is larger then tunneling with
simulataneous change of the vorticity, i.e. to  $\sim (x-iy)$ state, $t_L=J_b-J_a$. 
The difference is small since tunneling in the lowest Wannier state
$J_a\ll J_b$ but significant. The tunneling decreases the system energy thus the larger tunneling for the process
preserving vorticity will decrease the system energy more  and will be preferred. This observation allows
for including only right handed vortex in our single particle basis and omitting the left handed one.

\subsection{Single site anisotropy}

In our model we have assumed that the single site potential is isotropic, i.e. the two particle state produced
in the dipolar interactions has both the well defined energy and the relative angular momentum. 
In fact this is not strictly true. The single site potential cannot be approximated by a harmonic one
if fine details are to be studied. In a sqaure 2D lattice every site has four
neighbours and the square symmetry of the lattice influences the single site potential. The quartic terms
in the expansion of $V_0 (\sin^2(2\pi x)+\sin^2(2\pi y))$ potential are relevant.  
For this reason the two particle state, $|{\rm vortex}\rangle$ corresponding to $L_Z=2$ is not 
the eigenenergy state of single particle plus contact interaction on-site Hamiltonian. 
It is a superposition of three two-particles states of different energy instead. 
The fine structure results both from the anharmonicity and the anisotropy of the trapping potential. 
The anisotropy of the trap cannot be reduced
even for very high lattices. We checked that even for $V_0=100 E_r$ 
the energy splitting is significantly larger then dipolar interaction energy. Large magnetic moments, 
leading to larger dipolar energy could help to overcome this problem. 
Conservation of energy allows to tune independently only to the one of the three components 
of the $|{\rm vortex}\rangle$ state. Weak dipolar interactions 
resolve this fine structure of two-particle energy states.  
To observe the Einstein de Haas effect in optical lattices one should use the lattice geometry for
which the anisotropy due to the lattice symmetry is substantially reduced. To this end a 2D triangular
lattice with every site having 6 neighbours might be promissing. The other way out is to rotate
every lattice site around its axis similarly as in the experiment \cite{Gemelke}.

\section{Conclusions and outlook}
In this paper we studied the model Bose-Hubbard system
with two Wannier states in optical lattice.
We show that weak dipolar
interactions can be resonantly tuned to couple the ground Wannier
state to the excited one with higher orbital angular momentum. We
have studied a case of at most one particle per site on average. Even in
this case, we predict various novel phases of the system. 
The phase diagram of the system
significantly depends on the magnetic field. On the resonance 
we predict three distinct phases of the system: i) the Mott insulator
of superposition of ground and vortex states ii) the $a$-component dominated Mott 
superimposed with $b$-component dominated superfluid, iii) two superfluids in 
particular combination of both species.
We also discuss some limitations of our approach stressing that
harmonic approximation has to be used with caution when studying
orbital physics in optical lattices. 

Higher densities (more particles per site) are  more
favorable for dipolar transfer, the related physics will be discussed elsewhere.
It is worth noting that our results may be direcly related to the very recent
experiments, in which spin relaxation in an ultracold dipolar gas
in an optical lattice was observed in a presence of ultra low
magnetic field \cite{bruno, bruno2}. Although, so far, no vortices have
been found, we hope that the present work will help to identify
the regime of parameters,  in which generation of $P_x+iP_y$ superfluid and
appearance of novel quantum phases occurs.

\section{Acknowledgements}
The authors acknowledge discussions with M. Za{\l}uska-Kotur and J. Mostowski.
This paper was supported by the EU  STREP NAMEQUAM, IP AQUTE,
ERC Grant QUAGATUA, Spanish MINCIN (FIS2008-00784, QOIT),
Alexander von Humboldt Stiftung, Polish Ministry of
Science for 2009-2012 (J.Z.) and for 2009-2011 (M.G.) period.

\end{document}